\documentclass{webofc}
\usepackage[varg]{txfonts}   
\usepackage{graphicx,mathrsfs,amsmath,amssymb,amsopn,mathbbol}

\wocname{EPJ Web of Conferences}
\woctitle{Resonance Workshop at Catania}

\newcommand{\e}{\mathrm{e}}
\newcommand{\ii}{\mathrm{i}}
\newcommand{\erw}[1]{\ensuremath {\left \langle {#1} \right \rangle}}
\DeclareMathOperator{\im}{Im}

\newcommand{\MeV}{\ensuremath{\mathrm{MeV}}}
\newcommand{\GeV}{\ensuremath{\mathrm{GeV}}}

\newcommand{\dd}{\ensuremath{\mathrm{d}}}
\newcommand{\D}{\ensuremath{\mathrm{D}}}

\begin{document}

\title{Electromagnetic probes in heavy-ion collisions}
\subtitle{Messengers from the hot and dense phase}

\author{H.\ van Hees\inst{1,2}\fnsep\thanks{\email{hees@fias.uni-frankfurt.de}} \and
        J.\ Weil\inst{2}\fnsep\thanks{\email{weil@fias.uni-frankfurt.de}} \and
        S.\
        Endres\inst{2}\fnsep\thanks{\email{endres@th.physik.uni-frankfurt.de}} \and
        M.\ Bleicher\inst{2}\fnsep\thanks{\email{bleicher@th.physik.uni-frankfurt.de}}
}

\institute{Johann Wolfgang Goethe University Frankfurt, Institute for
  Theoretical Physics, Max-von-Laue-Str.\ 1, 60438 Frankfurt, Germany
\and
Frankfurt Institute of Advanced Studies, Ruth-Moufang-Str.\ 1, 60438 Frankfurt, Germany
}

\abstract{Due to their penetrating nature, electromagnetic probes, i.e.,
  lepton-antilepton pairs (dileptons) and photons are unique tools to
  gain insight into the nature of the hot and dense medium of
  strongly-interacting particles created in relativistic heavy-ion
  collisions, including hints to the nature of the restoration of chiral
  symmetry of QCD. Of particular interest are the spectral properties of
  the electromagnetic current-correlation function of these particles
  within the dense and/or hot medium. The related theoretical
  investigations of the in-medium properties of the involved particles
  in both the partonic and hadronic part of the QCD phase diagram
  underline the importance of a proper understanding of the properties
  of various hadron resonances in the medium.}

\maketitle

\section{Introduction}
\label{intro}

The transverse-momentum and invariant-mass spectra of the so-called
electromagnetic probes, i.e., dileptons ($\e^+ \e^-$ or $\mu^+ \mu^{-}$
pairs) and photons have been identified as interesting observables early
on \cite{Shuryak:1978ij}. Since they do not participate in the strong
interaction, they leave the hot and dense fireball created in
ultrarelativistic heavy-ion collisions nearly undisturbed by final-state
interactions and thus provide a direct insight into the spectral
properties of the electromagnetic current-current correlation function
in the medium during the entire evolution of the collision. For theory
this is also some challenge since an accurate description of the
invariant-mass spectra of dileptons and transverse-momentum spectra of
both dileptons and photons is needed, including as comprehensive a set
of sources as possible, reaching from the radiation from the very early
stage of the collision (Drell-Yan processes) over the emission from a
hot and dense partonic medium (Quark-Gluon Plasma, QGP) undergoing the
transition to a hot and dense hadron-resonance gas (which is close to
the chemical freeze-out of the medium), to the finally decoupled
hadronic state at thermal freeze-out.

In this paper we summarize the current status of our understanding of
both the spectral properties of the electromagnetic current-correlation
function, implying some insights about the nature of chiral-symmetry
restoration, and the description of the evolution of the hot and dense
partonic and hadronic fireball.

After a brief review in Sec.\ \ref{sect:dilepton-photon-sources} about
how to model the main thermal sources of electromagnetic probes in
heavy-ion collisions, in Sec.\ \ref{sec:fireball-evolution} we turn to a
short description of various models for the evolution of the hot and
dense fireball. Finally in Sec.\ \ref{sec:comparison-to-experiments} we
present the comparison of these models to recent data from the HADES
(GSI), NA60 (CERN SPS), and STAR (RHIC) collaborations and give and
outlook to what to expect from future experiments at FAIR and LHC.

\section{Sources of dileptons and photons}
\label{sect:dilepton-photon-sources}

Phenomenologically, at not too low collision energies the medium created
in heavy-ion collisions is well described as a collectively expanding
fluid in terms of hydrodynamical models
\cite{Teaney:2001av,Hirano:2002ds,Kolb:2003dz,Nonaka:2006yn}. This
implies that it is close to local thermal equilibrium over a large part
of its evolution. This suggests that models for the production of
electromagnetic probes, based on relativistic many-body quantum field
theory (for reviews cf.\ \cite{rw99,Rapp:2009yu}) are applicable.

Generally the radiation of photons and dileptons (``virtual photons'')
are given in terms of the retarded electromagnetic current-current
correlation function,
\begin{equation}
\label{1}
  \Pi_{\rm em}^{\mu \nu}(q_0,q)) = -\ii \int d^4x \ e^{iq\cdot x} \
  \Theta(x_0) \ \erw{ \left [\hat{j}_{\rm em}^\mu(x), \hat{j}_{\rm em}^\nu(0) \right]}. 
\end{equation}
For dileptons the invariant rate reads \cite{MT84,gale-kap90}
\begin{equation}
\label{2}
\frac{\dd N_{ll}}{\dd^4x \dd^4q} = -\frac{\alpha_{\rm em}^2}{\pi^3 M^2} \
       f_{\mathrm{B}}(q_0;T) \  \frac{1}{3} \ g_{\mu\nu} 
\ \im \Pi_{\rm em}^{\mu\nu} (M,q;\mu_B,T).  
\end{equation}
Here, $\alpha_{\text{em}} \simeq 1/137$ denotes the electromagnetic
coupling constant, $q_0$ and $q$ the energy and three-momentum of the
lepton pair in the restframe of the fluid cell at space-time position
$x$; $M$ is its invariant mass, $\mu_{\rm B}$ the baryon-chemical
potential, and $T$ the temperature of the medium.

In the following we briefly review some models used to describe the em.\
current correlation function of strongly interacting matter. In the
vacuum, the corresponding spectral function (\ref{1}) is empirically
given by the cross section for the reaction
$\e^++\e^- \rightarrow \text{hadrons}$ \cite{Agashe:2014kda}. In the
low-mass range $M \lesssim 1 \; \GeV$ it shows the low-lying
vector-meson resonances, $\rho$, $\omega$, and $\phi$ and in the
intermediate-mass range ($1 \; \GeV \lesssim M \lesssim 3 \; \GeV$) a
continuum. In pp and $AA$ collisions also the Dalitz decays of the
$\pi^0$ and $\eta$ Dalitz decays are prominent sources of dileptons.
 
\subsection{Thermal radiation from a QGP}

To lowest order the dominant dilepton-production process is the purely
electromagnetic annihilation of a quark-anti-quark pair to an
$\ell^+ \ell^-$ pair. The in-medium properties of the corresponding
production rate have been described using the leading-order
hard-thermal-loop formalism of thermal QCD \cite{bpy90} and more
recently at complete next-to-leading order, including a resummation
taking the Landau-Pomeranchuk-Migdal effect into account
\cite{Ghisoiu:2014mha} close to the light cone, $q^0=q$. Also results
from lattice-QCD calculations \cite{Ding:2010ga,Brandt:2012jc} are
available and have been extrapolated to finite three-momentum in
\cite{Rapp:2013nxa} and found in fair agreement with the microscopic QCD
description.

\subsection{Thermal radiation from a hadron-resonance gas}

At lower temperatures and densities, the in-medium current-correlation
functions have to be addressed using effective hadronic models. One
quite model-independent way is to use the corresponding empirical vacuum
spectral functions and low-density approximations to evaluate its medium
modifications or to employ so-called ``chiral reduction schemes''
\cite{Eletsky:2001bb,syz96,syz97}.

Beyond this leading-order low-density methods a detailed effective
hadronic model is necessary to evaluate reliable in-medium spectral
functions, including its three-momentum dependence. Starting from the
observation that the vector-meson dominance (VMD) model, i.e., the
assumption that the hadronic electromagnetic current is proportional to
the fields of the low-lying vector-meson resonances, $\rho$, $\omega$,
and $\phi$, provide a good description of the vacuum cross section,
various effective hadronic models have been developed, e.g.,
\cite{rw99,Teis:1996kx} and carefully fitted to the corresponding data
in ``elementary reactions'' like dilepton production in pp collisions or
photon-absorption data on nucleons and nuclei. Within such empirical
constraints the resulting dilepton-production rates based on the
evaluation of these models at finite temperature and/or density are in
satisfying agreement with each other.

According to these models medium modifications are due to two major
effects: (a) the medium modification of the pion cloud, i.e., the
dressing of, e.g., the pion loop in the $\rho$-meson self-energy diagram
due to hadronic interactions of the pions with the medium and (b) the
direct interaction of the vector mesons with mesons and baryons,
including a variety of resonance states. Both self-energy diagrams need
a proper dressing of the corresponding vertex functions to guarantee
electromagnetic gauge invariance. Besides the interactions with the
nucleon and $\Delta(1232)$ also heavier baryon resonances like the
N(1440), N(1520), N(1535), $\Delta(1700)$, etc. play an important
role. These interactions lead to a tremendous broadening of the
vector-meson spectral functions with quite moderate mass shifts. This
can be explained by the fact that any interaction process leads to an
increase of the imaginary part of the retarded self-energy
(``collisional broadening'') while its real part is lowered or raised
depending on the attractive or repulsive nature of the interaction. It
is important to note that also in heavy-ion collisions at the highest
available energies the interaction of the vector mesons with baryons are
the dominating source of these medium modifications, although the
net-baryon number (baryon-chemical potential) is low in this case. The
reason is that, due to the CP invariance of the strong interaction,
baryons and antibaryons add in the same way to the medium modifications
of the vector mesons, i.e., the relevant quantity is not the net-baryon
number density, $n_{\text{B}}-n_{\bar{\text{B}}}$ but the total one,
$n_{\text{B}}+n_{\bar{\text{B}}}$.

In addition, in the intermediate-mass region also contributions from
multi-pion processes (like $a_1,\omega$+$\pi$ reactions) are implemented
via low-density approximations, including vector-axial-vector mixing
\cite{Dey:1990ba}, using the empirical vacuum spectra of the vector- and
axial-vector current-correlation function from $\tau$-decay data
\cite{vanHees:2007th,vanHees:2006ng}.

\section{Fireball-evolution models}
\label{sec:fireball-evolution}

Having good models for the in-medium properties of the electromagnetic
current-correlation function at hand, as a second ingredient for a
realistic description of dilepton production in heavy-ion collisions
also a reliable model for the evolution of the partonic and hadronic
bulk medium is necessary, since the dilepton (and photon) spectra are
given by the integral of the corresponding rate (\ref{2}) over the
entire space-time four volume of the fireball during its whole
lifetime. While the integrated invariant-mass dilepton spectrum is a
Lorentz-invariant quantity, for the transverse-momentum spectra the
(radial) flow-velocity field of the medium becomes important due to the
corresponding Doppler-blue shift of the spectra of the irradiated
photons or dileptons
\cite{vanHees:2006ng,vanHees:2007th,Dusling:2006yv,Renk:2006qr}. The
description of the collective fireball dynamics reaches from simple
blastwave fireball parameterizations
\cite{vanHees:2006ng,vanHees:2007th,Renk:2006qr} over hydrodynamical
models
\cite{Dusling:2006yv,Vujanovic:2013jpa,Ryblewski:2015hea,Gale:2014dfa}
to detailed transport and ``hybrid'' transport-hydrodynamics models
\cite{Schenke:2005ry,Schmidt:2008hm,Bratkovskaya:2008bf,Linnyk:2011hz,Weil:2012ji,Santini:2011zw}.

One problem with the transport approach is the difficulty to implement
medium effects consistently. While the implementation of vacuum cross
sections and decay rates of hadrons into dileptons and photons is quite
straight forward, the medium modifications of such processes should, in
principle, be calculated within the same off-equilibrium setup as in the
corresponding kinetic description implemented in the transport models,
including the proper treatment of ``off-shell transport
prescriptions''. However this is out of reach with present
methods. Thus, more recently, a coarse-graining prescription has been
used to make the equilibrium-quantum-field-theory results for the
thermal dilepton/photon rates applicable within the underlying transport
model for the bulk evolution
\cite{Endres:2013cza,Endres:2013daa,Endres:2014xga,Endres:catania14}
(see also S.\ Endres's contribution to these proceedings
\cite{Endres:catania14}).

\section{Heavy-ion collisions at various energies}
\label{sec:comparison-to-experiments}

In this Section we present the comparison of some of the contemporary
models for dilepton production to recent data, covering an energy range
from $\sim 1 A \;\GeV$ (SIS at GSI) to top RHIC energy of
$200 A \;\GeV$.

\subsection{GSI and future FAIR experiments}

We start our comparison of the above summarized modeling of dileptons to
recent heavy-ion collision data with the measurements by the HADES
collaboration at the GSI SIS \cite{Agakishiev:2011vf,HADES:2011ab}. One
of the motivations to take data at low collision energies was the
verification of previous data in this collision-energy range by the DLS
collaboration, which have shown a large enhancement of the dilepton
yield in the low-mass region. The HADES measurement confirmed this
finding, providing a challenge for theory.
\begin{figure}[t]
\begin{minipage}{0.43 \linewidth}
\includegraphics[width=\linewidth]{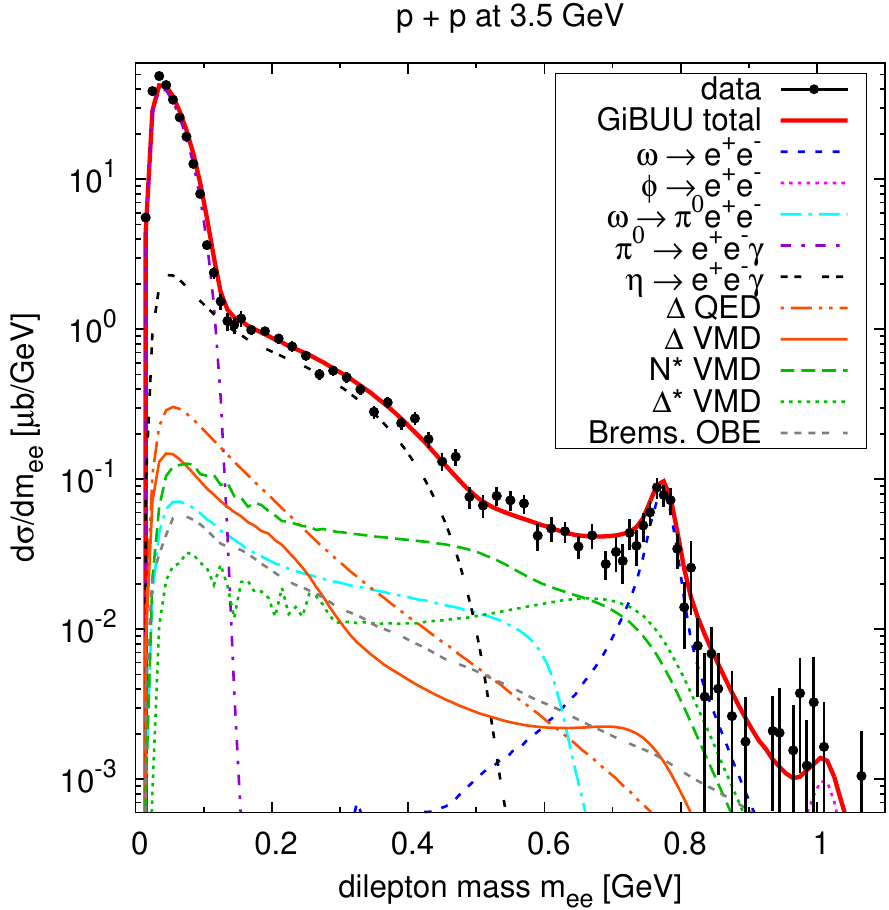}
\end{minipage} \hfill
\begin{minipage}{0.43 \linewidth}
\includegraphics[width=\linewidth]{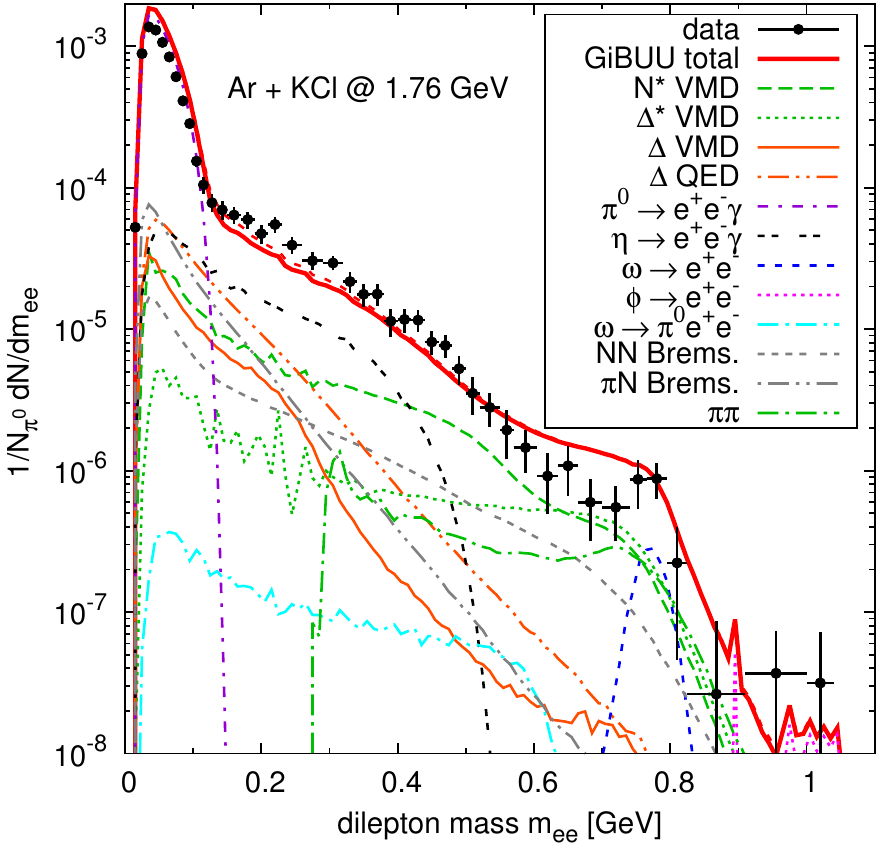}
\end{minipage}
\caption{Comparison of the dilepton invariant-mass spectrum from the
  GiBUU transport model in $3.5 \;\GeV$ pp (left) and $1.76 A\;\GeV$
  Ar-KCl collisions with data from the HADES collaboration
  \cite{HADES:2011ab,Agakishiev:2011vf}.}
\label{fig.1}
\end{figure}

As an example, in Fig.\ \ref{fig.1} we compare the results from the
GiBUU transport model
\cite{Buss:2011mx,Weil:2012ji,Weil:2012yg,Weil:2014lma,Weil:2014rea}
with the data from $3.5 \;\GeV$ pp and $1.76 A \;\GeV$ Ar+KCl collisions
from the HADES collaboration \cite{HADES:2011ab,Agakishiev:2011vf}. The
model implements an extension of the baryon-resonance model described in
\cite{Teis:1996kx} into the GiBUU transport simulation. The model
parameters (resonance masses, widths, and coupling constants) are fit to
the partial-wave analysis by Manley and Saleski
\cite{Manley:1992yb}. The coupling to the electromagnetic sector is
based on a strict VMD model, i.e., all dileptons are produced via an
intermediate light vector meson. This can be interpreted as a specific
model for the hadronic electromagnetic transition form factors in the
Dalitz-decay cross sections for the baryon resonances, including the
$\Delta(1232)$\footnote{Due to the p-wave nature of the $\rho N \Delta$-
  coupling, the impact on the hadronic properties of the $\Delta(1232)$
  resonance are negligible \cite{Weil:2014lma}.}. The sensitivity of the
transition form factor is illustrated in the left panel of Fig.\
\ref{fig.1}, where the VMD-form factor model is compared to a pure
electromagnetic (QED) coupling. The former provides a very satisfactory
description of dielectron production in elementary reactions. Only at
the lowest collision energies some discrepancies in comparison to the
HADES data in np collisions (from a quasi-free scattering analysis in dp
collisions) are found, which we trace back mostly to the difficulties to
fully describe and implement the bremsstrahlung contributions within the
transport-model approach \cite{Weil:2012ji}. The comparison of the model
with the Ar+KCl data (right panel of Fig.\ \ref{fig.1}) indicates the
possible influence of some medium modifications of the $\rho$ and
$\omega$ meson: A broadening of their spectra should decrease the
dilepton yield around the peak region,
$M \simeq m_{\rho} \simeq 770 \; \MeV$ and enhance it at lower
masses. The model also describes the measured transverse-mass ($m_t$)
and rapidity ($y$) spectra satisfactorily. A first comparison with the
Au-Au data, recently presented at the Quark Matter conference, also
indicates the need for medium modifications of the vector-meson spectral
functions.

Thus, we have also applied the coarse-graining approach to the UrQMD
transport model \cite{Endres:2015cg-hades} which enables us to use the
hadronic many-body theory for thermal dilepton rates by Rapp and Wambach
\cite{rw99b}, which is based on a similar hadron-resonance model as the
one used in the GiBUU transport model but evaluated at finite
temperature and baryon-chemical potential. Applying the medium modified
vector-meson spectral functions to the dilepton rates leads to an
excellent description of both the Ar+KCl (for the invariant-mass as well
as the $m_t$ and $y$ spectra) \cite{Endres:2015cg-hades}.

\subsection{SPS, RHIC}

\begin{figure}[t]
\begin{minipage}{0.47 \linewidth}
\includegraphics[width=\linewidth]{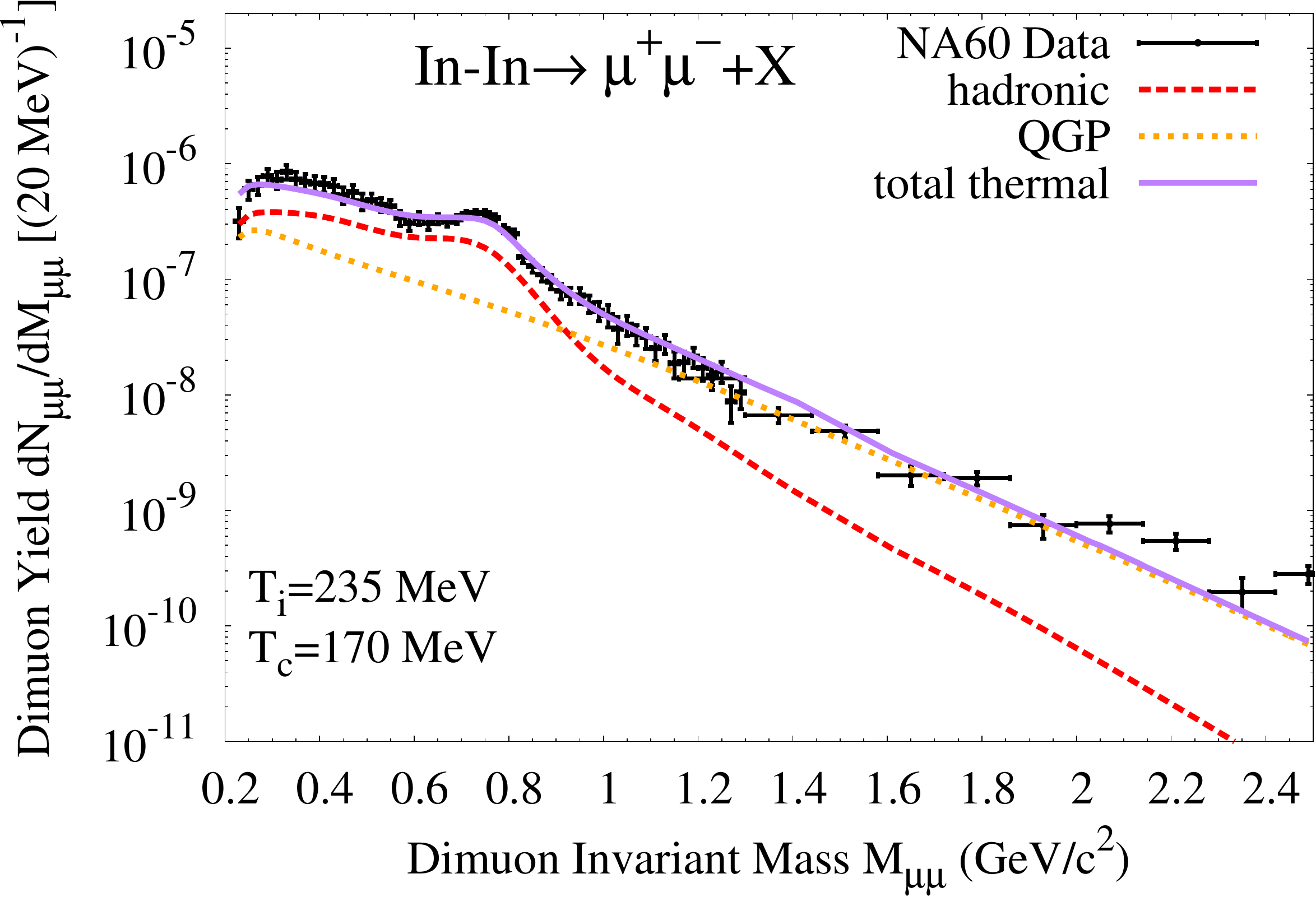}
\end{minipage} \hfill
\begin{minipage}{0.43 \linewidth}
\includegraphics[width=\linewidth]{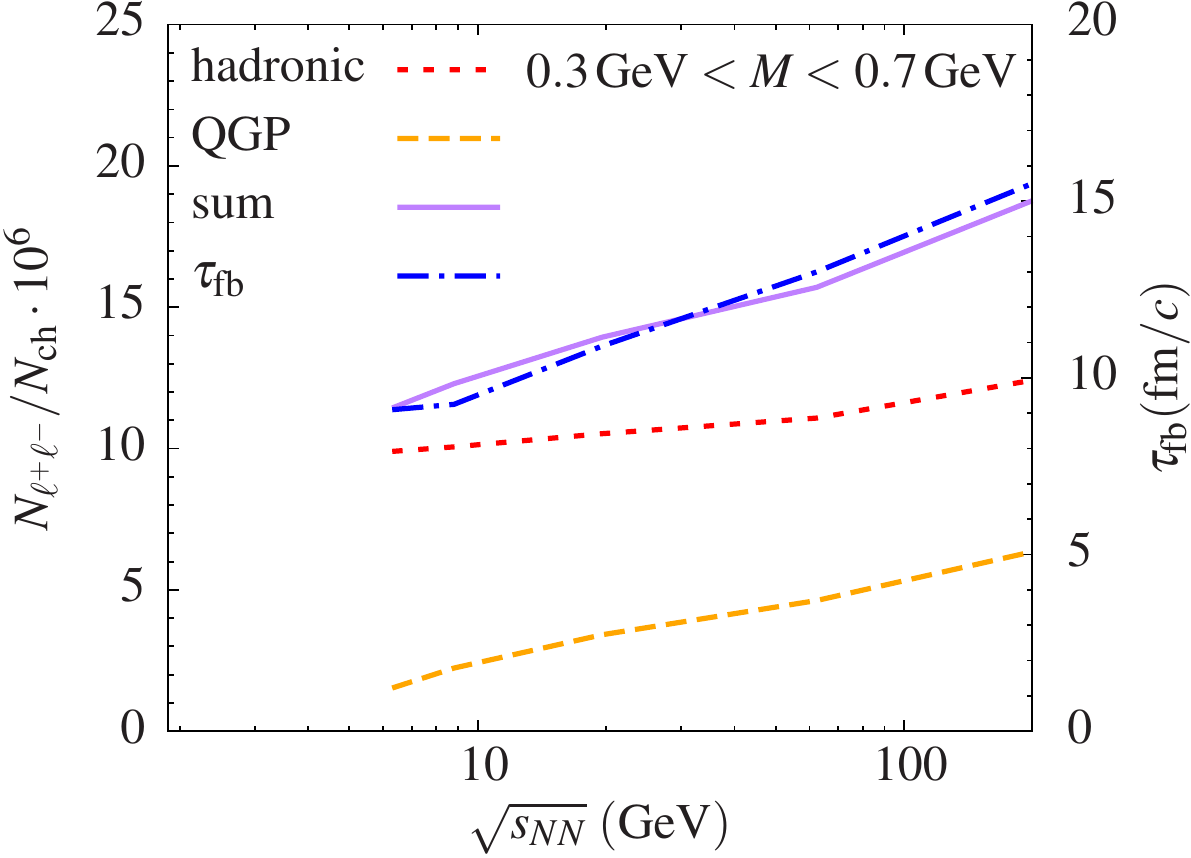}
\end{minipage}
\caption{Left: Invariant-mass spectrum with a simple blastwave thermal
  fireball model, using the medium-modified current correlation function
  of the Rapp-Wambach model in $158 A\;\GeV$ In+In collisions at the
  CERN SPS compared to acceptance corrected data from the NA60
  collaboration \cite{Specht:2010xu,Arnaldi:2008fw}. Right: Excitation
  function of the dilepton excess yield as a function of the center-mass
  beam energy in $AA$ ($A \simeq 200$) collisions.}
\label{fig.2}
\end{figure}
In the SPS energy regime we can compare the models to the high-precision
measurement of the dimuon excess spectrum by the NA60 collaboration in
$158 A\;\GeV$ In+In collisions. Based on data, here the Dalitz-decay
contributions at low masses, the yield from the decay of correlated
$\D \overline{\D}$ pairs as well as from the hard Drell-Yan processes in
the intermediate-mass region has been subtracted in a model-independent
way and made available as acceptance corrected mass spectra (also in a
comprehensive set of $p_t$ bins). Here, a considerable contribution from
the QGP phase as well as from multi-pion processes as described in Sec.\
\ref{sect:dilepton-photon-sources} is necessary to describe the data in
the intermediate-mass region. For the first time the accuracy of the
data has been sufficient to distinguish model predictions of the
dilepton spectra from hadronic-many-body theory, leading to a tremendous
broadening of the light vector meson's spectral functions with small
mass shifts, from those within a dropping-mass scenario based on
predictions from a particular realization (vector manifestation) of
chiral symmetry \cite{Brown:1995qt,Brown:1991kk,sas05,HS05}.

In addition, the slope of the acceptance-corrected invariant-mass
spectrum provides an excellent measurement of a space-time averaged
temperature of the dilepton source in this region, which is dominated by
emission from the QGP phase of the fireball evolution. Note that this
``invariant slope'' is unaffected from the Doppler blue-shift effect due
to radial flow, which is present in the corresponding slopes of the
transverse-mass spectra. The temperature extracted by a fit to the NA60
invariant-mass spectrum, where the non-thermal contributions from the
hard Drell-Yan process as well as from the decays of correlated
$\D\overline{\D}$ mesons have been subtracted,
$T_{M-\text{slope}} \simeq 205$-$230 \; \MeV$, is well above the
pseudo-critical temperature of the confinement-deconfinement and
chiral-symmetry (cross-over) phase transition of
$T_{\text{c}} \simeq 160 \; \MeV$.

Both the implementation of the medium-modified dilepton rates within the
coarse-graining approach with UrQMD
\cite{Endres:2014xga,Endres:catania14} and in a simple expanding
fireball model (cf. left panel of Fig.\ \ref{fig.2}) show an excellent
agreement with the NA60 data. This also holds for the $p_t$-differential
invariant-mass and the $p_t$ spectra measured by the NA60 collaboration
(see also S. Endres's contribution to these proceedings
\cite{Endres:catania14}). In addition, the same model for the in-medium
modified current-correlation function of strongly interacting matter has
been successfully employed to the dilepton measurement by the STAR
collaboration\footnote{The very large enhancement seen by the PHENIX
  collaboration \cite{Adare:2009qk} in the most central bin, cannot be
  described by the currently employed theoretical models.} at RHIC
\cite{Rapp:2013nxa} as well as the transverse-momentum spectra and
elliptic-flow parameter, $v_2(p_t)$, of ``direct photons'' at RHIC and
LHC \cite{vanHees:2011vb,vanHees:2014ida,Rapp:2014qha}. The surprisingly high
elliptic flow of the direct photons found at both RHIC and LHC underline
the importance of realistic models for photon production in the hadronic
phase, resulting in a dominant emission of photons from the phase of the
medium around the pseudo-critical temperature, showing a considerable
blue-shift effect in the effective slopes of the photon-$p_t$
spectra. Both findings indicate an early buildup of both radial and
elliptic flow. By comparison with hydrodynamic-model calculations for
the bulk evolution it has also been demonstrated that simple
blastwave-fireball parameterizations, constrained by hadronic freeze-out
spectra ($p_t$ and $v_2(p_t)$), are surprisingly reliable in describing
both the dilepton and photon spectra, based on the same in-medium model
for the electromagnetic current-correlation function. As shown above,
this also holds for the description of the fireball dynamics with
transport models as well as hybrid transport-hydro models.

\section{Conclusions and outlook}

These findings demonstrate that the electromagnetic probes (dileptons
and photons) in heavy-ion collisions are quite well understood in terms
of partonic and hadronic models for the medium modified electromagnetic
current-correlation function. In the low-mass region, the invariant-mass
spectra of dileptons show some sensitivity to the medium modifications
of vector mesons, underlining the importance of reliable effective
hadronic models, implementing the dynamics of a variety of meson and
baryon resonances, which have to be well-constrained by empirical input
on elementary cross sections. Here, the recent measurements of
pion-induced reactions by the HADES collaboration at the GSI SIS are
very promising for evaluating and improving these models in further
detail, particularly in the low collision-energy region.

The relative robustness of the electromagnetic probes to the details of
the bulk evolution, together with high-precision measurements of the
dilepton invariant-mass spectrum as well as $p_t$ spectra and $v_2(p_t)$
of direct photons (and also dileptons in the near future) can also help
to shed further light on the bulk-evolution properties. With simple
fireball parameterizations it is quite easy to reliably predict both the
invariant effective slope (and thus a space-time averaged true
temperature, unaffected by blue-shift effects due to radial flow) and
the life-time of the fireball, which is the only parameter to be
adjusted to the overall dilepton or photon yield, as soon as the
partonic and hadronic models for the in-medium production rates and the
bulk-medium evolution model are fixed.

This enables us to make predictions of various ``excitation functions''
like the invariant-slope temperature, the dilepton excess yield in the
low-mass region, and the fireball lifetime (cf. the right panel of Fig.\
\ref{fig.2}) as a function of the center-mass beam energy, which is
addressed currently in the beam-energy-scan program at RHIC and at the
future CBM experiment at the FAIR project. The here shown excitation
functions are based on an cross-over equation of state for all beam
energies, constrained by lattice-QCD calculations
\cite{Rapp:2014hha}. Since at least the invariant-slope temperature is
quite sensitive to the employed equation of state, the hope is to find
indications for a change from a cross-over phase transition at higher
beam energies (low baryon-chemical potential) to a first-order phase
transition (higher baryon-chemical potentials) or maybe even the
existence of a critical point in the QCD phase diagram. Via the
invariant-slope temperature at least an indication for the transition
from a fireball evolution starting in a partonic phase to one which is
entirely of hadronic nature should be possible. In turn these
alterations of the phase-transition behavior should also be reflected in
the total lifetime of the fireball (``critical slowing down'' close to a
critical point) and the directly related overall yield of em.\ probes.


\end{document}